\documentclass[english,aps,prb,preprint]{revtex4}
\usepackage{graphicx}
\usepackage{amssymb}
\usepackage{textcomp}  
\usepackage{hyperref}
\hypersetup{colorlinks=true,linkcolor=blue,citecolor=red,urlcolor=black}

\begin{document}

\title{Controlling photonic structures using optical forces }

\author{Gustavo S. Wiederhecker}

\author{Long Chen}

\author{Alexander Gondarenko}

\author{Michal Lipson}

\affiliation{School of Electrical and Computer Engineering, Cornell University,
Ithaca, New York 14853, USA.}

\begin{abstract}
The downscaling of optical systems to the micro and nano-scale results in very compliant systems with nanogram-scale masses, which renders them susceptible to optical forces. Here we show a specially designed resonant structure for enabling efficient static control of the optical response with relatively weak repulsive and attractive optical forces. Using attractive gradient optical forces we demonstrate a static mechanical deformation of up to 20 nanometers in the resonator structure. This deformation is enough to shift the optical resonances by roughly 80 optical linewidths.
\end{abstract}
\maketitle

\section{Introduction}

The use of optical forces to manipulate small objects is an important
subject with applications ranging from manipulation of living cells
by optical tweezers \cite{Ashkin:1987lk}  to optical cooling in atomic physics \cite{Hansch:1975wj}. More
recently, the downscaling of optical systems to the micro and nano-scale
resulted in very compliant systems with nanogram-scale masses, rendering
them susceptible to optical forces \cite{Antonoyiannakis:1999ng,Ng:2005qc,Povinelli:2005qp,Rakich:2007cj}. In fact, optical forces have
been exploited to demonstrate chaotic quivering of microcavities \cite{Carmon:2007wf},
optical cooling of mechanical modes \cite{Kippenberg:2007sy,Kippenberg:2008wu,Schliesser:2008df}, actuation of a tapered-fiber
waveguide \cite{Eichenfield:2007ys}, and excitation of the mechanical modes of silicon nano-beams
\cite{Li:2008fb}. The challenge in controllably manipulating the optical response
of photonic structures using optical forces is that large optical
forces are required, in order to induce appreciable changes in the
structures geometries. Here we show a specially designed resonant
structure for enabling efficient static control of the optical response
with relatively weak optical forces. We demonstrate a static mechanical
deformation of up to 20 nanometers in the geometry of a silicon nitride
structure, using three milliwatts of optical power. Because of the optical
response sensitivity to this deformation, such optically induced static
displacement introduces resonance shifts spanning 80 times the intrinsic resonance
linewidth. 

\begin{figure*}
\includegraphics[width=16cm]{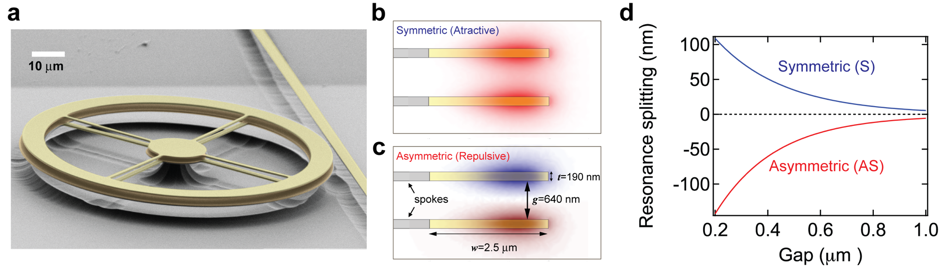}

\caption{\label{fig1}\textbf{Design of a resonant cavity sensitive to optical gradient forces}. (a) Scanning electron micrograph of two vertically stacked ring cavities. (b,c) Cross-section of the ring region showing the calculated electric field profile for  the symmetric (b) and antisymmetric (c) optical modes supported by the cavity. The ring thickness is $t=190$ nm, width $w=2.5$ \textmu m  and inter-ring gap $g\approx 640$ nm. The symmetric mode induces attractive optical forces while the antisymmetric induces repulsive forces. (d) Typical simulated resonant wavelength splitting of the S and AS optical modes as the  gap between the rings is changed.}

\end{figure*}

\section{Results}

The photonic structure is composed of two coupled ring resonators
with an optical response sensitive to small changes in the distance between the rings. As shown in Fig. \hyperref[fig1]{\ref{fig1}a}, the structure is composed of a
pair of vertically stacked silicon-nitride (Si$_3$N$_4$) rings held by very
thin spokes and a pedestal. The spokes are designed to reduce the
mechanical stiffness of the cavity, increasing the sensitivity to the
optical forces between the rings, and at the same time induce only
negligible scattering for light circulating near the outside edge
of the rings\cite{Anetsberger:2008om} (see Methods for fabrication details). Due to the relatively small refractive index of Si$_3$N$_4$
($n\approx2.0$), strong optical coupling can occur for
relatively large gaps between the top and the bottom resonator. As a result of this coupling,
the transverse mode profile splits into symmetric (S, Fig. \hyperref[fig1]{\ref{fig1}b}) and
antisymmetric (AS, Fig. \hyperref[fig1]{\ref{fig1}c}) combinations, leading to two distinct
resonant frequencies. Figure \hyperref[fig1]{\ref{fig1}d} shows an example of the resonance
wavelength splitting (centered at 1493 nm) for the transverse electric
(TE) mode of a 30 \textmu m diameter ring, obtained from numerical
simulations (see Supplementary Information). One can see that the splitting depends exponentially
on the separation gap. Such steep gap dependence of the resonant frequencies
on the separation translates into strong gradient optical forces between
the rings \cite{Povinelli:2005qp,Rakich:2007cj,Lin:2009mz} (see Supplementary Information for details). Here we show that light coupled to the symmetric or antisymmetric optical modes can induce
attractive or repulsive optical forces. 

\begin{figure*}
\includegraphics[width=17cm]{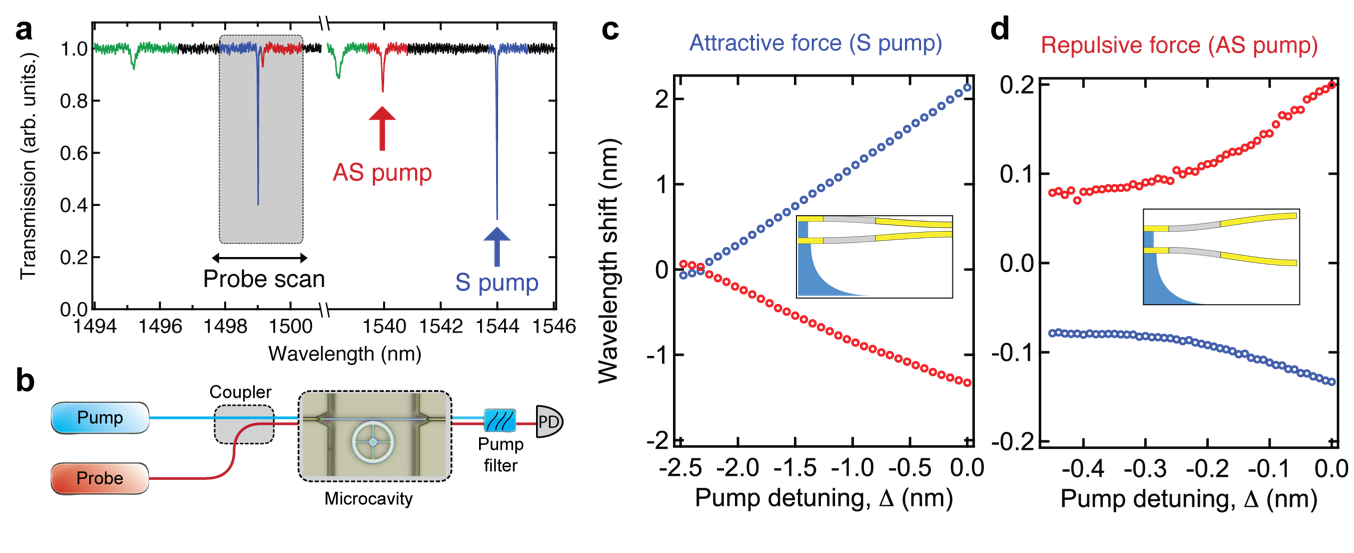}

\caption{\label{fig2}\textbf{Demonstration of gradient force control of cavity resonances}. (a) Optical transmission of the fabricated device shown in Fig. \hyperref[fig1]{\ref{fig1}a}. The different colors in the transmission spectrum indicates different optical modes. The blue one is the fundamental symmetric TE mode ($TE_1^{+}$), the red one is the fundamental antisymmetric (AS) mode ( $TE_1^{-}$), and the green one is the second order symmetric mode ($TE_2^{+}$). (b) Schematic of the experimental setup. The pump and probe laser light are coupled and launched into the device, residual pump is filtered out and detected by the photodiode (PD). (c) Measured resonance splitting when the pump laser is tuned towards a symmetric resonance (blue arrow) inducing attractive forces. The blue curve represents the probed S mode while the red curve represents the probed AS mode . (d)  Same as in (c) but with the pump laser is tuned to a antisymmetric resonance (red arrow) inducing repulsive forces. The cartoons inset on parts (c) and (d) illustrates the optical force induced deformation (out of scale) of the structure.}

\end{figure*}

Both symmetric and antisymmetric optical modes can be observed in the transmission spectrum of the fabricated structure, indicating the coupling between the two cavities. In Fig. \hyperref[fig2]{\ref{fig2}a} we show the optical
transmission for input light with transverse electric polarization (TE)
for the 30 \textmu m diameter ring shown in Fig. \hyperref[fig1]{\ref{fig1}a}. For this device, the coupling gap between the bus waveguide and the rings is 650 nm. Each mode can be identified by comparing their measured free spectral ranges
with numerical simulations (see Supplementary Information for details). The typical loaded optical quality factors
for the fundamental $TE_1^{+}$ and $TE^{-}$ resonances were $Q=6.8\times10^4$ (finesse $\mathcal{F}=720$) and $Q=2.1\times10^4$ ($\mathcal{F}=200$), respectively. We attribute this difference to the asymmetry between the top and the bottom ring geometry and material properties (see Methods and Supplementary Information for details).

We demonstrate experimentally the control of the photonic structure\textquoteright{}s
response using both attractive and repulsive forces, i.e. the tuning of the resonance splitting with optical
forces.  In contrast to other tuning methods based on thermal effect or free-carrier injection, this approach allows the resonances to be moved to either shorter or longer wavelengths. As illustrated in the schematic of Fig. \hyperref[fig2]{\ref{fig2}b}, we use a pump
laser to induce the optical force and a weak probe laser to read-out
the cavity response (see methods for details). Attractive and repulsive forces are obtained by tuning the control laser towards a symmetric (blue arrow in Fig. \hyperref[fig2]{\ref{fig2}a}) or antisymmetric optical resonance (red arrow in Fig. \hyperref[fig2]{\ref{fig2}a}). For each detuning wavelength between the pump laser and the respective cavity resonance, the probe laser wavelength is swept across a pair of S and AS resonances (shaded region in Fig. \hyperref[fig2]{\ref{fig2}a}) to read out the resonance shift. The probed wavelength shifts (relative to $\lambda=1499.1$ nm) for both S excitation (attractive forces) and AS excitation (repulsive forces) are plotted in Fig. \hyperref[fig2]{\ref{fig2}c,d}. In Fig. \hyperref[fig2]{\ref{fig2}c}, while the AS  resonance (red circles) is blue shifted, the S resonance (blue curve) is red shifted. This opposing behavior can only occur if the gap between the two rings is reducing as the pump laser approaches the symmetric cavity resonance. In Fig. \hyperref[fig2]{\ref{fig2}d} we show the wavelength splitting measured when the AS mode is excited. In contrast to the S excitation shown in Fig. \hyperref[fig2]{\ref{fig2}c}, the AS resonance is red shifted while the S resonance is blue shifted. As we discuss next, such opposite resonance shift uniquely demonstrates that optical gradient force is the dominant actuation effect in our cavity. Moreover, the ability to induce both attractive and repulsive optical forces in a microcavity structure allows one to shape the optical potentials to explore novel effects and functionalities in optomechanical devices, such as self-alignment and or cavity trapping\cite{Rakich:2007cj}. 

Due to optical absorption, thermal effects are also expected to induce resonance shifting.  As the cavity temperature increases, the refractive index of Si$_3$N$_4$ will increase due to thermo-optic effect. Also, because of the thermal expansion, the cavity structure may be deformed inducing a thermo-mechanical effect. The thermo-optic effect causes the resonances to red shift, creating thus an asymmetry between S and AS splitting (see Supplementary Information for details). Such thermo-optic asymmetry can be noticed in Fig. \hyperref[fig2]{\ref{fig2}c} where the S mode red shift is larger than the AS mode blue-shift. In  \hyperref[fig2]{\ref{fig2}c}, due to the repulsive optical force, the asymmetry is reversed and the S mode blue shift is smaller than the AS mode red shift. The thermo-mechanical effect however can actually deform the mechanical structure. This deformation can either increase or decrease the gap between the rings.  A raise in temperature however will always deform  induce the same thermo-mechanical deformation, regardless of the optical mode that is providing the heat source. This is in clear contrast with the measurements shown in Fig. \hyperref[fig2]{\ref{fig2}c,d}. To infer the magnitude of this effect in our structure, we performed full numerical simulations of coupled thermo-mechanical problem using the finite element method (see Supplementary Information). The results show that even a very large temperature increase of $\Delta T=70$ K would  change the gap between the rings by only 100 pm. Such small contribution occurs because the thermal expansion is mainly in the radial direction, leaving the gap between the rings almost unchanged. 

From the measured the net resonance shift for both S and AS modes, the thermo-optic ($\Delta\lambda_{to}$) and the optomechanical ($\Delta\lambda_{om}$) contributions to the net resonance shift can be isolated. This is achieved by considering the measured group index of the optical modes, the thermo-optic coefficient of SiN$_{x}$ and the optomechanical tuning efficiency ($k_{om}=\partial\lambda/\partial g$) (see Methods and Supplementary Information). The optomechanical
tuning efficiencies for S and AS resonances are obtained from the
numerical simulation of the cavity's optical response.
For a 640 nm inter-ring gap they have the numerical
values $k_{om}^{+}=-0.07$ and $k_{om}^{-}=0.1$. The maximum optomechanical and thermal shift, obtained for attractive forces, is calculated to be $\Delta\lambda_{om}^+\approx1.5$ nm  ($\approx80$ linewidths) for the S mode and $\Delta\lambda_{om}^-\approx-2$ nm ($\approx30$ linewidths) for the AS modes, the thermo-optic contribution is similar for both modes and is calculated to be $\Delta\lambda_{to}^\pm\approx0.6$. 

\begin{figure*}
\includegraphics[width=16cm]{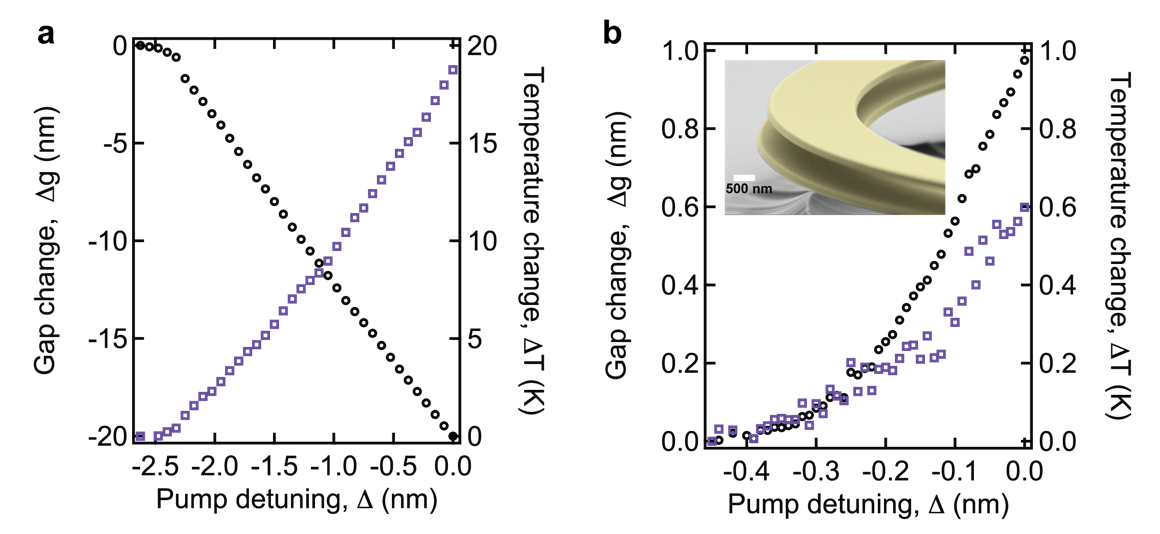}

\caption{\label{fig3}\textbf{Inter-cavity gap and temperature change}. Change in the gap between the resonators (black curve, left scale) induced by the optical force and temperature increase due to optical absorption (purple curve, right scale). (a) Symmetric pump (attractive forces). (b) Antisymmetric pump (repulsive forces). The inset in part (b) shows a zoomed in scanning electron microscope image of the cavity, the scale bar is 500 nm.}

\end{figure*}

The change in the gap between the two rings
induced by the optical force can also be estimated from the splitting curves
shown in Fig. \hyperref[fig2]{\ref{fig2}c,d} and the optomechanical tuning efficiencies. The maximum gap change is found to be $\Delta g^+=-20$ nm for attractive force and $\Delta g^-=1$ nm for repulsive forces. This difference are mostly due to the different quality factors of these modes. The tuning of this gap with respect to the pump laser detuning is shown in Fig. \ref{fig3}a,b. Using a spring constant of $k=1.2$ N/m, calculated
from numerical simulations of the cavity's mechanical
response, the optical force on each ring can be estimated from the
usual linear relation $F^+=k\Delta g^+/2\approx7.5$ nN. This value is in the same order of magnitude with the value expected from the cavity energy $U$ and optomechanical tuning efficiency \cite{Povinelli:2005qp,Rakich:2007cj}, $F^+=(U^+/\lambda) k_{om}^+\approx12$ nN. For repulsive forces this yields $F^-=k\Delta g^-/2\approx600$ pN whereas the predicted force is $F^-=(U^-/\lambda) k_{om}^-\approx350$ pN (see Supplementary Information for details).  

We verify that the impact of undesirable optically induced
mechanical oscillations in our structure is not detrimental by measuring the
dynamic optical response (see Methods and Supplementary Information). Due to its higher optical $Q$ such oscillations are  more relevant to the symmetric pump excitation. Because the optical quality factor ($Q\approx6\times10^4$) of this mode is not very high and the mechanical quality factor ($Q_m\approx2$) is extremely low, the threshold for regenerative mechanical oscillations\cite{Kippenberg:2007sy} of the fundamental mechanical mode is estimated to be $P_{th}\approx65$ mW and was not achieved in our experiment. The thermal brownian motion of the cavity mechanical modes however can cause fluctuations in the gap between the rings. At thermal equilibrium,  this amplitude can be calculated using the equipartition theorem to be $\sqrt{\left<\delta x^2\right>}\approx60$ pm. This corresponds to relative wavelength shift $\left<\Delta \lambda\right>/\delta\lambda= k_{om}\sqrt{\left<\delta x^2\right>}/\delta\lambda\approx 20\%$. In our cavity however a giant optical spring effect\cite{Eichenfield:2009cr} is observed due the small effective mass ($m_{eff}\approx 85$ pg) and relative large optomechanical tuning efficiency,  . The frequency of the  fundamental mechanical mode is observed to increased by a 14-fold in our structure, this corresponds to a 200-fold increase in the effective spring constant (see Supplementary Information). Because the thermal fluctuation scales with the inverse of the spring constant, the optical spring should reduce the thermal fluctuation to $\sqrt{\left<\delta x^2\right>}\approx4$ pm, corresponding to a relative wavelength fluctuation $\left<\Delta \lambda\right>/\delta\lambda= k_{om}\sqrt{\left<\delta x^2\right>}/\delta\lambda\approx 1\%$, where $\delta\lambda$ is the resonance linewidth. The relative power fluctuation of the transmitted pump laser was measured to be $\left<\Delta P\right>/P\approx 0.05$ at small pump detunings. This corresponds to relative wavelength shift $\left<\Delta \lambda\right>/\delta\lambda\approx 7\%$. These could  be reduced
by targeting an even lower mechanical quality factor cavities using, for instance,
a pressurized environment or specific spokes design \cite{Anetsberger:2008om}.

\section{Conclusions}
We show that significant optical force actuation is achievable by exploring interacting
optical microcavities. The particular cavity design explored in this
work provides high mechanical sensitivity to such forces. For instance,
within the optical and mechanical quality factors explored here, the static optical force is dominant over both  thermal effects and optomechanical oscillations. Demonstration of  both attractive and repulsive forces is an important step towards enabling recently
proposed functionalities for optomechanical devices, such as self-aligning
and optical corralling behavior \cite{Rakich:2007cj}. These advances on the static actuation
of photonic microcavity structures using optical forces should enable
future micro-optomechanical systems (MOMS) with novel and distinct
functionalities.

\begin{acknowledgments}

This work was supported in part by the National Science Foundation under grant 00446571. The authors also acknowledge partial support by Cornell University's Center for Nanoscale Systems. This work was performed in part at the Cornell Nano-Scale Science \& Technology Facility (a member of the National Nanofabrication Users Network) which is supported by National Science Foundation, its users, Cornell University and Industrial users.
\end{acknowledgments}

\appendix
 \setcounter{figure}{0}
\renewcommand{\theequation}{S-\arabic{equation}}
\renewcommand{\thefigure}{S-\arabic{figure}}

\section{Methods}

\subsection{Device fabrication}

The two layers (190 nm thick each) of stoichiometric Si$_3$N$_4$ were deposited
using low-pressure chemical vapor deposition (LPCVD) while the 300
nm thick SiO$_{2}$ layer was deposited by plasma-enhanced chemical vapor
deposition (PECVD). The underlying substrate is a 4 \textmu m
SiO$_{2}$ formed by thermal oxidation of a silicon wafer. Note that because the top Si$_3$N$_4$ layer is deposited over a PECVD SiO$_2$, the surface roughness of the top layer is expected to be larger than the bottom layer, decreasing its optical quality compared to the bottom  Si$_3$N$_4$ layer. In order to have a single coupling waveguide, we perform two lithography steps. First circular pads are defined in the position where the cavities will be patterned. The upper Si$_3$N$_4$ layer is then etched, except on the circular pads region. In the second lithography step, both the wheel pattern and the waveguides are defined, however, the wheels are defined on top of the previously defined circular pads. Only then the intermediate SiO$_{2}$ and bottom Si$_3$N$_4$ layer are etched. As a result the waveguides will have a single Si$_3$N$_4$ layer whereas the wheels will have two layers.
After depositing a 1 \textmu m thick protective SiO$_{2}$
cladding using PECVD, we use optical lithography to pattern the spun
photoresist with a rectangular window around the resonators. In order
to release the structure, the device is immersed in buffered hydrofluoric
acid for an isotropic etch of the SiO$_{2}$ in the window region. To avoid
stiction of the stacked disks, the device is subsequently dried using
a critical point dryer. Due to the internal tensile stress ($\sigma_0=1$ GPa) of the Si$_3$N$_4$ films, there is a noticeable bending of the rings in the vertical direction ( see the SEM shown Fig. \hyperref[fig2]{\ref{fig1}a}). Due
to such bending, the actual gap between the resonators is larger than the sacrificial SiO$_2$ layer thickness. It is estimated
from the SEM image to be $g=(640\pm50)$ nm.

\subsection{Experimental set-up}

The set-up used to actuate and probe the cavity is illustrated in
Fig. \hyperref[fig2]{\ref{fig2}b}. It consisted of two tuneable external cavity diode lasers
combined using a 3 dB directional coupler. The laser light was coupled in and out
of the chip using a pair of lensed optical fibers. Due to the high insertion 
loss of our devices ($12.0\pm0.5$ dB), the pump laser was
pre-amplified using an erbium doped fiber amplifier (not shown in
the schematic of Fig. \hyperref[fig2]{\ref{fig2}b}). To minimize coupling fluctuations when the pump power is varied, we choose to keep the pump laser power constant and change the detuning from pump laser wavelength $\lambda_p$ with respect to desired cavity resonances. In both case the laser approached the resonance from smaller wavelengths (blue detuning). The light collected from the bus waveguide is split in two parts. One part is filtered with an interferometric filter (center wavelength 1550 nm, bandwidth 10 nm) to reject the residual pump light. The remaining probe light is detected and provide the raw optical transmission spectra used to obtain the curves shown in Fig. \hyperref[fig2]{\ref{fig2}c,d} , see Supplementary Information for the raw transmission data. The second part is directly detected using a New Focus 1811 (125 MHz bandwidth) photodetector. Its photocurrent allows us to measure the dynamical response of the microcavity and measure the RF spectrum of the transmitted pump laser (see Supplementary Information).

\subsection{Coupled pump power calibration}

The input pump power ($P_{in}$) launched in the coupling waveguides is estimated from the input and output loss of our devices ($=12\pm0.5$ dB each). To ensure that both input and output losses were the same we have calibrated them based on the nonlinear wavelength shift obtained by launching light from opposite sides of our device. 

\subsection{Optical modes}
Three optical modes can be identified in the measured transmission spectrum. As indicated in Fig.  \hyperref[fig2]{\ref{fig2}a}, we identified them as the fundamental symmetric mode in blue ( $TE_1^+$  (blue), fundamental antisymmetric in red ($TE_1^-$), and the second order symmetric in green ($TE_2^+$). This identification is obtained by  comparing the measured calculated free-spectral ranges (FSR) with the simulated ones (see Supplementary Information).  Another evidence that this identification is correct comes from the actual wavelength shift experienced by each of these modes. Although we only show the wavelength shift for the S ($TE_1^+$) and AS $TE_1^-$ modes on Fig. \hyperref[fig2]{\ref{fig2}}, we extended the probe laser scanning window to measure the $TE_2^+$ mode resonance shift. As shown in the Supplementary Information, the $TE_2^+$ experiences a red-shift for attractive forces (symmetric pump) as it should be for a symmetric mode.

\subsection{Gap change calculation}

The thermo-mechanical effect is negligible in our structure, therefore the measured resonance shifts are  due to a combination of thermo-optic and optomechanical effect.  The measured shift for S ($+$) and AS ($-$) modes will be given by $ \Delta\lambda^{\pm}  = k_{om}^{\pm}\Delta g+ k_{to}^ {\pm} \Delta T$. Since both $\Delta\lambda^+$ and $\Delta\lambda^-$ can be measured in our cavities and both $k_{om}$ and $k_{to}$ can be estimated, the system of two equations can be solved for the gap ($\Delta g$) and temperature change ($\Delta T$). The corresponding contribution to the net resonance shift can easily calculated as $\Delta\lambda_{om}^\pm=k_{om}^{\pm}\Delta g$ and $\Delta\lambda_{to}^\pm=k_{to}^{\pm}\Delta T$. 

\subsection{Dynamical response}

The temporal response of the resonator was investigated by analyzing
the time-domain response of the transmitted pump laser using a 1 GHz real-time oscilloscope and a RF spectrum analyzer. Under the
maximum power explored here (3 mW of coupled pump power) the measured
relative oscillation amplitude in the pump transmission was $\delta P/P\approx0.05$.
By fitting a lorentzian lineshape to the pump resonance, one can map
this power fluctuation into a drift of the resonance wavelength. For
small pump detuning ($<5$ pm) the estimated wavelength
drift is approximately 1.6 pm, which corresponds roughly to 7\% of the pump
resonance linewidth. Inspecting the RF spectrum of the transmitted
pump laser we also identified the mechanical resonances responsible
for the observed oscillations (see Supplementary Information). The mechanical
quality factors for the fundamental mechanical mode was $Q_m=2$.


\section{\label{sec:sup}Supplementary Information}

\subsection{Optical Force and mechanical actuation}

The gradient optical force between the two rings can be predicted by considering the optical potential energy for the symmetric and antisymmetric optical modes\cite{Povinelli:2005qp,Rakich:2007cj},
\begin{equation}
F=-\frac{\partial U}{\partial g}=\frac{U}{\lambda} k_{om},
\label{eq:optical_force}
\end{equation}
where $U$ is the intra-cavity field energy, $\lambda$ is the wavelength of light and $k_{om}=\partial \lambda/\partial g$ is the optomechanical wavelength tuning efficiency.  This latter quantity is related to the usual optomechanical frequency tuning parameter ($g_{om}$) as $g_{om}=(\omega/\lambda) k_{om}$. Symmetric modes will have negative $k_{om}$, leading to attractive forces ($F<0$) whereas antisymmetric modes will have a positive one, leading to repulsive forces ($F>0$). To calculate the optical force in terms of the experimental parameters we write the  cavity energy as $U=(1-T)P_{in}Q_i\lambda/(2\pi c)$, where $T$ is the transmission extinction ratio, $Q_i$ is the intrinsic quality factor, $P_{in}$ is the input pump power coupled to the bus waveguide, and $c$ is the speed of light. The optical force can be related to motion of each ring using Hooke's law $F=k(\Delta y)$, where $k$ is the mechanical spring constant and $\Delta y=\Delta g/2$. This predicts a gap change between the rings given by
 \begin{equation}
\Delta g=\frac{(1-T)P_{in}Q_i k_{om}}{\pi c k} .
\label{eq:predicted_deltag}
\end{equation}
In Fig. \ref{figA:raw_split}a,b the optical resonances excited by the pump laser to induce both attractive and repulsive forces are shown. The pump power in both cases was $P_{in}^{max}=(3\pm0.4)$ mW. Using the experimental parameters for the S resonance, $T=0.31$, $Q_i=8.73\times10^4$, and $k_{om}=-(7\pm2)\times10^{-2}$ obtained from numerical modelling (see Fig. \ref{figA:split_model}) yields a maximum gap change $\Delta g^+=-(12\pm 3)$ nm for the attractive pump.  Similar calculations applies for the repulsive pump, the parameters in this case are $T=0.83$, $Q_i=2.26\times10^4$ and $k_{om}=0.1\pm0.02$, yielding a maximum gap change of $\Delta g^-=(0.6\pm0.1)$ nm for the repulsive pump. 

These values estimated for the optical gradient force have the same order of magnitude of the data presented on Fig. \ref{fig3}a,b. The maximum measured gap change shown in Fig. \ref{fig3} were  $\Delta g^+\approx-20$ nm   and $\Delta g^-\approx1$ nm.  We attribute these differences to the uncertainty in the simulated optomechanical tuning efficiency as we discuss below.

\subsection{Optical modes and optomechanical tuning efficiency}

\subsubsection{Optical modes identification}
 The cavity optical fields are assumed to have azimuthal symmetry, i.e. $\vec E(r,z,\phi)=\vec E(r,z)\exp(im\phi)$ so that it's optical modes could be calculated for each azimuthal number $m$ in a 2-D computational domain.   For simplicity, we have also approximated the cavity structure by two identical rings which are perfectly aligned to each other, i.e. the gap is constant over the whole ring region. Using numerical simulations of this cavity structure we identify three guided optical modes in 1500-1600 nm wavelength range. They are the fundamental symmetric ($TE_1^+$, Fig. \ref{figA:split_model}b),  the fundamental  antisymmetric ($TE_1^-$, Fig. \ref{figA:split_model}c), and the second order symmetric TE mode ($TE_2^+$). The measured free spectral range (FSR) is shown in \ref{figA:split_model}a for each mode. The blue curve is for the $TE_1^+$, the green curve for the $TE_2^+$ while the red curve represent the $TE_1^-$. In the inset of \ref{figA:split_model}a we show the corresponding calculated FSR in the same scale. Although there is small mismatch between the actual values, the trend observed in the experimental curves are reproduced quite well. 
 
 Another strong experimental evidence that supports our identification stems from the measured wavelength splitting for the $TE_2^+$ mode. Although this is not shown in the main text, its wavelength shift can be observed in Fig. \ref{figA:raw_split}b, where the probe scanning range includes the $TE_2^+$ resonance around $\lambda=1495$ nm. As expected for a symmetric mode, it experiences a red shift when the optical force induces attraction between the rings.

 \subsubsection{Optomechanical tuning efficiency}
\begin{figure}
\includegraphics[width=16cm]{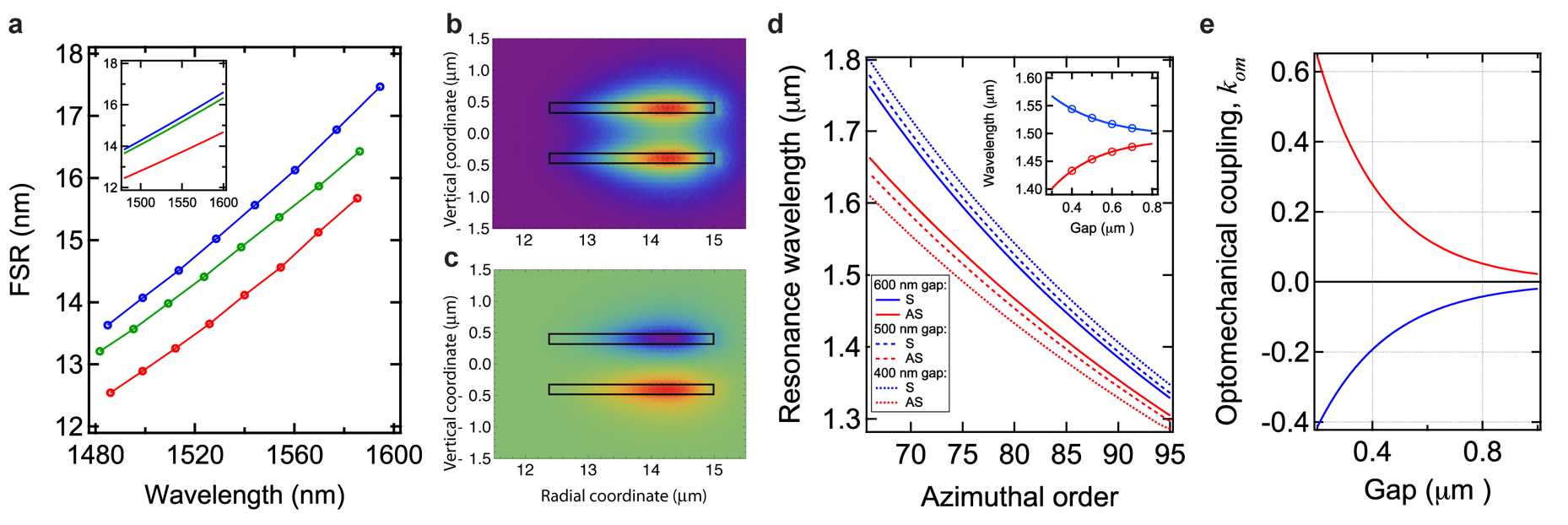}

\caption{\label{figA:split_model}\textbf{Optical modes and optomechanical tuning efficiency}. Simulated optical response of the double-wheel resonator. (a) Measured free spectral range (FSR) for the modes highlighted in Fig. \ref{fig2}, the blue represents the $TE_1^+$, the red represents the $TE_1^-$ and the green stands for the $TE_2^+$. The inset shows the respective FSR obtained from simulation (b,c) Symmetric optical mode  and antisymmetric optical mode for a gap of $g=600$ nm. (d) Optical resonance wavelength for different azimuthal order ($m$) for both S (blue) and AS (red) modes. The different line styles correspond to different gaps between the two rings.  The circles in the inset shows the resonance wavelength (azimuthal number $m=81$) for the different gaps. The solid lines represent the fit described in the text.  (e) Optomechanical coupling ($k_{om}=\partial\lambda/\partial g$) obtained from the fitted curves shown in the inset of part (d).}

\end{figure}

\begin{figure*}
\includegraphics[width=16cm]{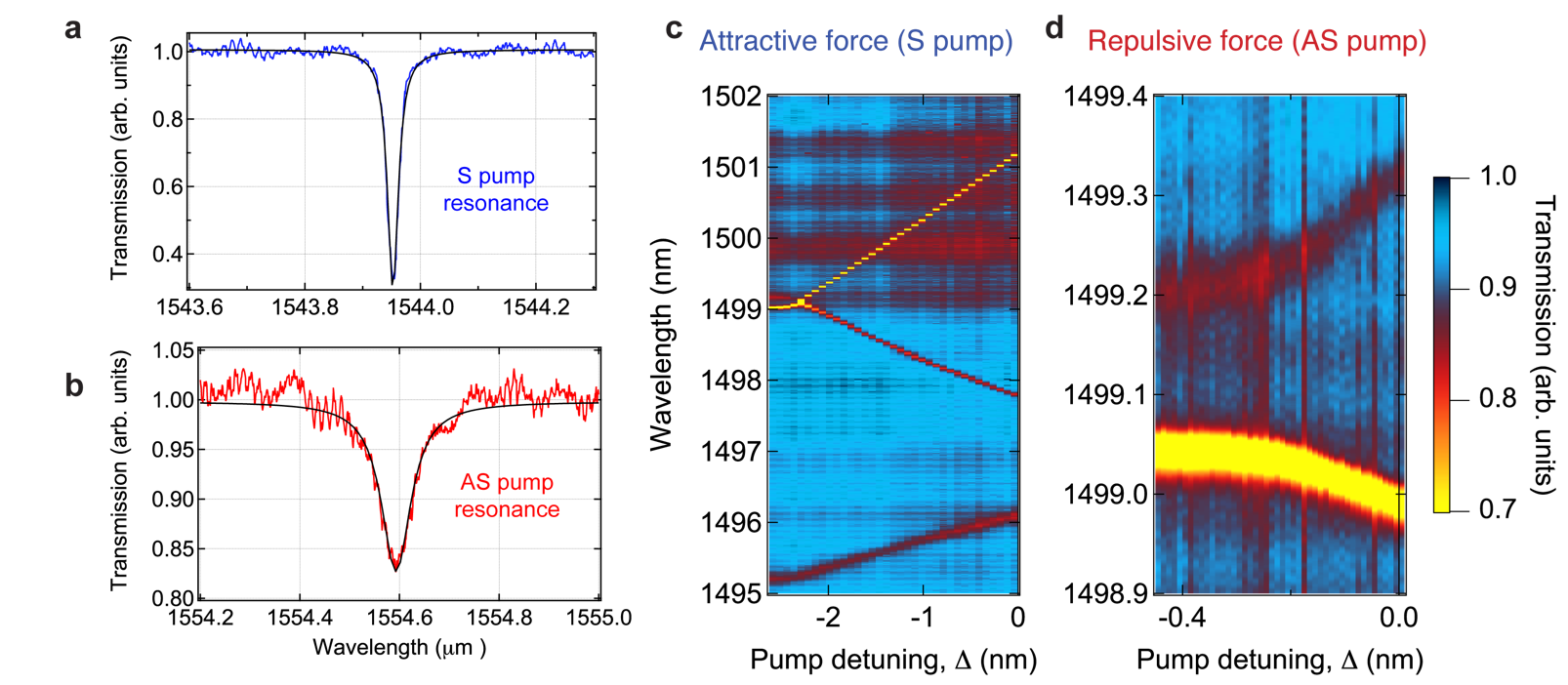}

\caption{\label{figA:raw_split}\textbf{Optical force induced resonance splitting}.(a,b) Pump resonances used for inducing attractive and repulsive forces. (c,d) Measured optical transmission (false color) around the probe resonances when the pump laser wavelength is swept over the pump resonances shown in (a,b).}

\end{figure*}

The optomechanical coupling parameter  $g_{om}$ is obtained by calculating the resonant mode frequencies for different gaps. In Fig. \ref{figA:split_model}c, the resonance wavelengths for both the fundamental transverse electric (TE) S and AS optical modes are shown for different gaps between the rings. The inset in Fig. \ref{figA:split_model}c shows an exponential fit to the the resonant wavelength (for azimuthal order $m=81$) in the form $\lambda^\pm=w_0^\pm+w_1^\pm exp(-g w_2^\pm)$. The fit parameters for the S mode were $w_0^+=1493.2  \pm 0.2$ nm, $w_1^+=232.3 \pm 0.3$, and $w_2^+=(3.82 \pm 0.01)\times 10^{-3}$. For the AS modes, the fit parameters were $w_0^-=1493.2  \pm 0.2$ nm, $w_1^-=336.1 \pm 0.3$, and $w_2^-=(4.35 \pm 0.01)\times 10^{-3}$. 

The gap dependent optomechanical tuning efficiency can be easily calculated from this fit as $k_{om}^\pm(g)=\partial\lambda^\pm/\partial g= -w_1^\pm w_2^\pm \exp (-g w_2^\pm)$. Fig. \ref{figA:split_model}d shows such parameters for both S (blue) and AS (red) modes. In our device, the gap is estimated from the SEM micrograph to be $g=640\pm50$ nm, therefore $k_{om}=(7\pm2)\times10^{-2}$. Note that the error estimated here for $k_{om}$ is purely due to the uncertainty in the gap retrieved from the SEM measurement. However we have verified numerically, and there is also recent experimental evidence \cite{Lin:2009mz}, that small asymmetries between the top and bottom rings may cause  optical mode coupling between the fundamental AS mode and the second order S mode. This coupling will affect the resonance splitting for the the AS mode and due to the anti-crossing nature of such splitting \cite{Lin:2009mz}, the optomechanical tuning efficiency for the AS mode could be reduced.  This would contribute to additional uncertainty in the calculated gap change. Any asymmetry also will induce a different power distribution among the top and bottom rings for S and AS modes. Because the top Si$_3$N$_4$ layer in our device should have a higher scattering loss (see fabrication details in the Methods section), this can lead to considerable difference between the top and bottom resonators.


\subsection{Gap change calculation}

When the pump laser is coupled  to the cavity, the optical force is not the only effect to contribute to the resonance shift. Other nonlinear effects may also contribute to shift the optical resonances. In silicon nitride, possible contributions could arise from thermal and electronic (Kerr) nonlinearity. For the pump power levels used, however, contributions from Kerr effect can be neglected. Thermal effects though are expected to contribute in two different ways.

The simplest one is through the positive thermo-optic effect in which a temperature variation ($\Delta T$) induces a proportional change in the material refractive index ( $\delta n = \alpha \Delta T$, $\alpha=4\times10^{-5}$ K$^{-1}$)   for Si$_3$N$_4$). As a result of positive thermo-optic effect there is always a red-shift of the cavity resonances when any light is absorbed in the rings. The impact of the thermo-optic effect on the resonance position can be easily quantified using the usual relation $\Delta\lambda_{to}^{\pm}  = k_{to}^ {\pm}\Delta T$, where $k_{to}= \alpha \left(\lambda/ n_g\right)^{\pm}  $ is the thermo-optic parameter,   $n_g$ is the group index and the superscript $\pm$ stands for either the symmetric ($+$) or antisymmetric optical mode ($-$).

The other thermal contribution can arise due to thermo-mechanical effect. Because our structure is suspended and the rings are thermally isolated due to the thin spokes, the temperature gradient in the structure can actually induce some deformation of the cavity structure. In contrast to the thermo-optic effect, which can only red-shift the optical resonances, such deformations could also change the gap between the rings and therefore shift symmetric and antisymmetric resonances to different directions. For a given cavity however,  the thermo-mechanical deformation will either increase the coupling between the rings or reduce it. Therefore, if this effect was dominant, when the pump laser is coupled to a symmetric or antisymmetric mode it would induce resonance shifts in the same direction.  This is in contrast with our observations shown in Fig. \hyperref[fig2]{\ref{fig2}c,d}. Such behavior can only be accounted for if attractive and repulsive optical forces are dominant in our structure. Therefore thermo-mechanical effect should be small in our structure. To confirm that we have also performed thermo-mechanical simulations of our structure, showing that the gap change is negligible for the observed temperature shifts (see below).

The total resonance shift for S ($\lambda^+$) and AS ($\lambda^-$) modes is a result of thermo-optic and optical force effect, $ \Delta\lambda^{\pm}  = k_{om}^{\pm}\Delta g+ k_{to}^ {\pm} \Delta T$, where $k_{om}$ and $k_{to}$ are, respectively, the optomechanical and thermo-optic wavelength tuning efficiencies. The optomechanical wavelength tuning efficiency is related to the usual optomechanical coupling rates as $ k_{om}\equiv\partial \lambda/\partial g =g_{om}c/\lambda^2$,  where $g_{om}$ is the usual optomechanical coupling parameter. Since both $\Delta\lambda^+$ and $\Delta\lambda^-$ are measured one can calculate both the gap ($\Delta g$) and temperature change ($\Delta T$) as (see Fig. \ref{fig3})
\begin{eqnarray}
\Delta g = \frac{-k_{to}^{-} (\Delta\lambda^{+} )+ k_{to}^{+}( \Delta \lambda^{-})}  {k_{om}^{+}k_{to}^{-}   -  k_{om}^{-} k_{to}^{+}} \label{eq:delta_G}\\
\Delta T = \frac{k_{om}^{-} ( \Delta\lambda^{+} )- k_{om}^{+}( \Delta \lambda^{-})}  {k_{om}^{+}k_{to}^{-}   +  k_{om}^{-} k_{to}^{+}}
\label{eq:delta_T}
\end{eqnarray}

\begin{figure*}
\includegraphics[width=12cm]{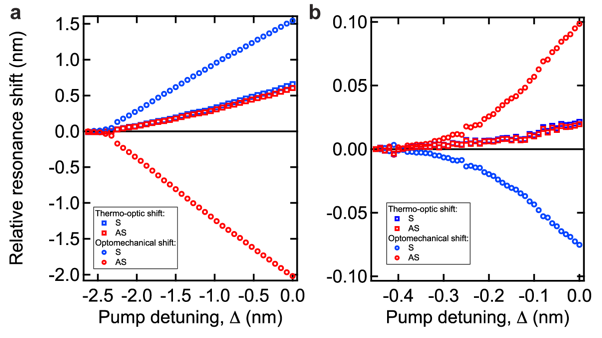}

\caption{\label{figA:relative_shift}\textbf{Relative resonances shift}.  Individual contributions to the resonance shift as calculated using the temperature and gap change given by Eqs. \ref{eq:delta_G},\ref{eq:delta_T}. (a) Attractive pump; the blue curves correspond to the S modes while the red curves correspond to the AS modes. The circles represent the optomechanical shift ($\Delta \lambda_{om}=k_{om}\Delta g$) while the squares represent the thermo-optic shift ($\Delta \lambda_{to}=k_{to}\Delta T$). (b) Same as in (a) for the repulsive pump.}

\end{figure*}
\subsection{Thermal properties}
The steady-state temperature change in the cavity is governed by the heat equation \cite{Carmon:2004bq},
\begin{equation}
\Delta T=\Gamma_{abs} R_{th} U,
\label{eq:heat}
\end{equation}
where $\Gamma_{abs}$ is the cavity optical absorption rate and $R_{th}$ is the cavity thermal resistance. Before diffusing to the substrate, heat must diffuse through the spokes and the pedestal. The thermal resistance of each of these elements will be given by given by $R_{th}=l/(k_{th}A)$, where $l$ is the length,  $k_{th}$ is the material thermal conductivity, and $A$ is the cross-section area. The spokes in our structure have $l=9.5$ \textmu m  and $A=0.1$  \textmu m$^2$. Using $k_{th}=20 $ W/m/K for  Si$_3$N$_4$ yields $R_{th}^{spk}\approx5\times10^{6}$ K/W. Similar calculations yields $R_{th}^{pds}\approx5\times10^4$ for the SiO$_2$ pedestal. The total thermal resistance is then $R_{th}\approx R_{th}^{pds}+R_{th}^{spk}/8\approx0.7\times10^6$ K/W, where the numerical factor 1/8 stems from the total of eight spokes on the two rings. The optical absorption rate $\Gamma_{abs}$ is estimated from the absorption limited quality factor as $Q_{abs}=\omega/\Gamma_{abs}$. A typical value for the absorption limited quality factor of Si$_3$N$_4$ is $Q_{abs}\approx 5\times10^6$ and therefore $\Gamma_{abs}/2\pi\approx36$ MHz. Using these values we can estimate the temperature raise in the cavity for both S($+$)and AS($-$) pumping using Eq. \ref{eq:delta_T}. The maximum cavity energy achieved in our experiments were $U^+\approx150$ fJ and  $U^-=5$ fJ, therefore the temperature should increase approximately by $\Delta T^+\approx25$ K and $\Delta T^-\approx0.9$ K. These number are within the same order of magnitude of the temperatures calculated from the measured resonance shift using eq. \ref{eq:delta_T} (see Fig. \ref{fig3})

\subsubsection{Thermo-mechanical effect}

\begin{figure*}
\includegraphics[width=14cm]{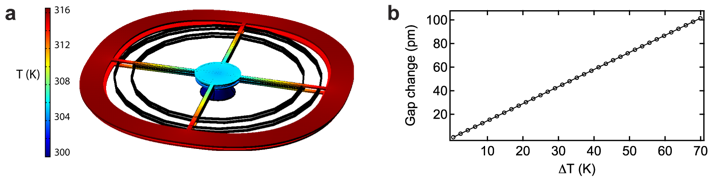}

\caption{\label{figA_thermal_comsol}\textbf{Thermo-mechanical effect}. (a) Cavity deformation and temperature map due to thermo-mechanical effect for total heating power $P_{heat}=35$\textmu W. The black-solid lines represent the cold cavity geometry. (b) Calculated gap change as a function of the ring temperature due to thermo-mechanical effect. }

\end{figure*}

Although the observation of both attractive and repulsive optical forces suggest that thermo-mechanical effects should be of secondary importance in our structure, we have also performed thermo-mechanical numerical simulations using the finite element method. Since the largest temperature variation occurs for the symmetric pump excitation (attractive forces), we used a heat source  distributed over the outer circumference of each ring with heating power given by $P_{heat}=\Gamma_{abs}U^+\approx35$ \textmu W. The steady-state temperature distribution in the cavity is shown in Fig. \ref{figA_thermal_comsol}a. Temperatures gradients between the rings and the pedestal of 16 K can be noticed. This value allows to calculate a thermal resistance of $R_{th}=\Delta T/P_{heat}\approx0.5\times10^6$ K/W, in reasonable agreement with the estimations above. In Fig. \ref{figA_thermal_comsol}c we show the simulated cavity deformation together with the temperature map (false color scale). The solid black lines represent the cold cavity geometry. Note that most of the thermal-induced displacement occurs in the radial direction and therefore weakly perturbs the optical modes. Varying the heat power $P_{heat}$ we calculate the change in the gap between the rings as a function of ring temperature. The result is shown in Fig. \ref{figA_thermal_comsol}d, where a maximum change of $\delta g\approx100$ pm is observed for a temperature change of $\Delta T\approx 70$ K. These results confirm that thermo-mechanical effect is negligible in our structure.

\subsection{Dynamical response}

Although this work is focused on the static mechanical response of the double rings, there is also a inherent dynamical behavior. Even in the absence of light, or at very low power levels, the coupling to the thermal bath at a finite temperature $T$ leads to position fluctuations following equipartition of energy, 

\begin{equation}
\left<\delta x^2\right>=\frac{k_BT}{m_{eff}\Omega_m^2}.
\label{eq:thermalx}
\end{equation}
For the lowest frequency mechanical modes of our structure ($\Omega/2\pi=600$ KHz, $Q_m=2$, $\Gamma_m=\Omega/Q_m=1.9$ MHz,  $m_{eff}=k/\Omega_m^2=85$ pg), Eq. \ref{eq:thermalx} yields $\sqrt{\left<\delta x^2\right>}\approx60$ pm. This corresponds to a optical resonance fluctuation of $\left<\Delta \lambda\right>= k_{om}\sqrt{\left<\delta x^2\right>}\approx 4$ pm, this corresponds to a relative shift with respect to cavity linewidth of $\delta\lambda/\left<\Delta/\lambda\right>\approx18\%$

Due to the dynamic back-action of the mechanical motion one should expect dynamical behavior beyond the simple thermal motion described above \cite{Kippenberg:2007sy,Pinard:1999pi}. It is known that the  initially static optical force can change the  dynamical properties of the resonator. This is known as optical spring effect \cite{Eichenfield:2009cr,Kippenberg:2007sy,Lin:2009mz,Pinard:1999pi}. Such optical spring has a component which is  in-phase with the mechanical spring which shifts the resonant mechanical frequency $\Omega_m$. There is  also a quadrature-component which leads to changes in the mechanical damping parameter $\Gamma_m$, resulting in either damping or amplification of the thermally excited mechanical modes. As a result of such dynamical back-action, the mechanical resonant frequency and damping will be modified depending on the pump laser frequency to cavity frequency detuning ($\Delta_0=\omega_p-\omega_0$) as \cite{Eichenfield:2009cr,Kippenberg:2007sy}
\begin{eqnarray}
\Omega_m'^2=\Omega_m^2 + \left(\frac{2Ug_{om}^2}{\omega m_{eff}}\right)   \frac{\Delta_0}{\Delta_0^2+(\Gamma/2)^2}\\
\label{eq:spring_effect}
\gamma_m'=\gamma_m - \left(\frac{2Ug_{om}^2\Gamma }{\omega m_{eff}}\right)   \frac{\Delta_0}{\left[\Delta_0^2+(\Gamma/2)^2\right]^2}
\end{eqnarray}

\begin{figure*}
\includegraphics[width=16cm]{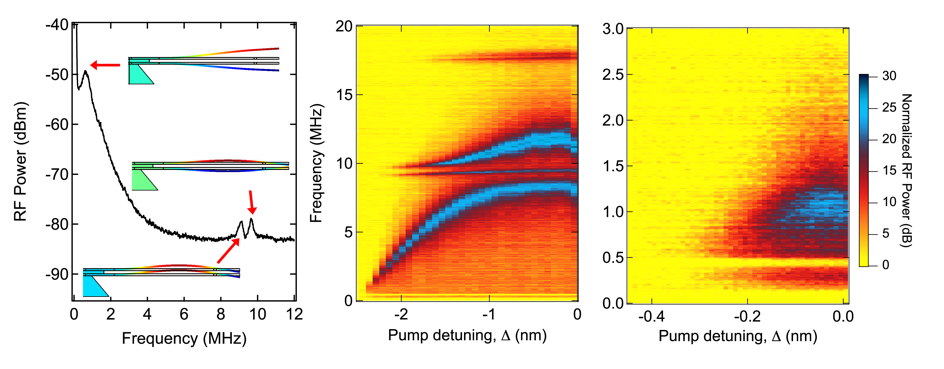}

\caption{\label{figA_optical_spring}\textbf{Giant optical spring effect}. RF spectrum  of the transmitted pump laser showing the mechanical resonances of the cavity (a) Low pump power ($P\approx10$ \textmu W) RF spectrum. The insets indicates the mechanical mode profile obtained from numerical simulation. (b) Evolution of the pump RF spectrum as a function of detuning with respect to the symmetric resonance at ($\lambda\approx1544$ nm). The pump power was $P\approx3$ mW. The fundamental mechanical mode exhibits giant optical spring effect and has its frequency shifted from 600 KHz to 8 MHz. (c) Same as in (b) but for the antisymmetric resonance at ($\lambda\approx1555$ nm).}

\end{figure*}

Due to the optical spring effect, the effective mechanical frequency of the resonator changes and therefore the thermal motion amplitude given by Eq. \ref{eq:thermalx} is drastically reduced.  In Fig. \ref{figA_optical_spring} we show the measured RF spectrum of the transmitted pump laser as a function of the detuning between the laser wavelength and the cold cavity resonant wavelength. A giant spring effect can be noticed, where the fundamental oscillation mode has it frequency shifted from 600 KHz to 8 MHz. This frequency shift corresponds to almost 200-fold increase in the spring constant. Due such large optical spring effect, the device is more stable against thermal noise when the pump laser is on. Using Eq. \ref{eq:thermalx} with a new equilibrium temperature $T=320$ K (due to optical absorption), this results in displacement fluctuation of only $\sqrt{\left<\delta x^2\right>}\approx4$ pm. The corresponding relative wavelength fluctuation reduces to  $\delta\lambda/\left<\Delta/\lambda\right>\approx1\%$

In our experiment, because we pump our cavity resonances at blue-detuning, the mechanical damping rate is reduced as the pump laser loads the cavity. Therefore there is amplification of the thermal motion. Such optically amplified  vibrations will become detrimental for static actuation when the back-action gain overcome the intrinsic mechanical dissipations \cite{Kippenberg:2007sy}. Setting $\Gamma_m'=0$ in Eq. \ref{eq:spring_effect} leads to the threshold power for the parametric instability,

\begin{equation}
P_{th}=\frac{\Omega_m m_{eff} \omega^4}{4 g_{om}^2 Q_m Q^3}
\label{eq:threshold}
\end{equation}

Using the experimental values described above, the estimated threshold from Eq. \ref{eq:threshold} for the symmetric mode excitation is $P_{th}\approx65$ mW. Within the power levels used in our experiments this threshold was not achieved.

\bibliographystyle{naturebst}
\bibliography{grad_force_references}

\end{document}